\documentclass[reprint, prd, nofootinbib, superscriptaddress]{revtex4-1}
\usepackage{graphicx}
\usepackage{amsmath}
\usepackage{xcolor}
\usepackage{graphics}
\usepackage[normalem]{ulem}
\begin{document}


\title{Revisiting the stability of strange-dwarf stars and strange planets}
\author{Victor P. {\sc Gon\c{c}alves}}
\email{barros@ufpel.edu.br}
\affiliation{Institut f\"ur Theoretische Physik, Westf\"alische Wilhelms-Universit\"at M\"unster,
  Wilhelm-Klemm-Stra\ss e 9, D-48149 M\"unster, Germany}
\affiliation{Institute of Modern Physics, Chinese Academy of Sciences,
  Lanzhou 730000, China}
\affiliation{Institute of Physics and Mathematics, Federal University of Pelotas, \\
  Postal Code 354,  96010-900, Pelotas, RS, Brazil}

\author{Jos\'e C. {\sc Jim{\'e}nez}}
\email{josec.jimenez91@gmail.com}
\affiliation{Instituto de F\'{\i}sica, Universidade de S\~ao Paulo,\\
Rua do Mat\~ao 1371, 05508--090 S\~ao Paulo-SP, Brazil}

\author{Lucas {\sc Lazzari}}
\email{lucas.lazzari@ufpel.edu.br}
\affiliation{Institute of Physics and Mathematics, Federal University of Pelotas, \\
  Postal Code 354,  96010-900, Pelotas, RS, Brazil}
\begin{abstract}
  The  dynamical stability of strange-dwarf hybrid stars and strange planets, constituted by strange-quark-matter cores and dilute-nuclear-matter crusts, is revisited by analyzing the fundamental mode eigenfrequencies of the radial oscillation equations with boundary conditions for slow (rapid) conversions originating at the density-discontinuous interface characterizing {extremely large (small) microscopic timescales compared to the oscillation periods}.
  For the hadronic crust we used an analytic fit of the BPS results matched to the massless MIT bag model.
  For the rapid case, our calculations indicate that the zero mode is the so-called {\it reaction mode} whose frequency is a complex number, thus ruling out the existence of strange dwarfs (planets) in nature.
  On the other hand, slow conversions {still provide a sizeable stability window which, interestingly, also reproduces the} Glendenning-Kettner-Weber {results}.
  The robustness of our findings is demonstrated for different transition densities and using an equation of state from perturbative QCD for the ultra-dense core.
\end{abstract}
\maketitle
\section{Introduction}
Compact stars are formed at the end of stellar evolution when the gravitational pull is strong enough to melt atoms keeping only some fundamental fermions as basic constituents providing enough pressure to balance the imminent gravitational collapse, e.g. electrons in white dwarfs giving maximal masses around ${1.4}{M_{\odot}}$ and neutrons in neutron stars (NS) with maximal masses around ${0.7}{M_{\odot}}$, both in the ideal limits \cite{Glendenning:2000}.
Currently, the measurements of NS masses being around ${2}{M_{\odot}}$~\cite{Linares:2018ppq,Miller:2019cac,Cromartie:2019kug,Riley:2019yda} may indicate the presence of exotic phases at their cores as, e.g. quark matter, which can arise from the {non-perturbative} confinement/deconfinement transition {of quantum chromodynamics (QCD)} expected to occur at intermediate baryon densities~\cite{Annala:2019puf}.
However, the description of the NS cores is still a theme of intense debate and the same mass limit can be achieved for purely hadronic NSs~\cite{Lonardoni:2014bwa}.

A directly related challenging problem is the characterization of the interiors of all the observed white dwarfs (WD) reported in Ref.~\cite{Kurban:2022}, whose masses and radii do not coincide with those expected for usual WDs. Such objects could be better explained by the so-called strange dwarfs (SD), which are exotic -- yet hypothetical -- stellar objects composed by a dilute nuclear crust (essentially WD matter) and a strange quark matter (SQM) core, being {\it strange} as proposed by Bodmer, Witten and Terazawa~\cite{Glendenning:2000} {(see, e.g., Refs. \cite{Glendenning:1994zb,DiClemente:2022ktz} for proposals of their stellar formation).
Almost three decades ago, the authors of Refs.~\cite{Glendenning:1994sp,Glendenning:1994zb}, using a massless MIT bag model for the core and the Baym-Pethick-Sutherland (BPS) \cite{Baym:1971pw} equation of state (EoS) for the hadronic mantle, have argued that SDs are dynamically stable and might exist in nature when subject to infinitesimal radial perturbations for reasonable values of the transitional density.
Nevertheless, Ref.~\cite{Alford:2017vca} suggested that these numerical calculations are wrong and the usual Bardeen-Thorne-Meltzer (BTM) criterion~\cite{Bardeen:1966} still works for these hybrid stars\footnote{See, e.g. Ref.~\cite{Vartanyan:2012zz} (and references therein) where only the static stability condition was used to characterize stable strange dwarfs.}.
For better numerical performance, they made a fitting on the BPS equation of state as well as an hyperbolic smoothing on the discontinuous sector of the whole hybrid EoS. Notice that their findings were obtained assuming the neutron drip point as the only transitional density.

More recently, Ref.~\cite{DiClemente:2022ktz} implemented non-trivial boundary conditions for the Lagrangian variables, which account for slow and rapid conversions of fluid elements at the quark-hadron interface when subjected to radial oscillations~\cite{Haensel:1989wax,Pereira:2017rmp}.
Using this formalism, the BPS EoS and a slightly modified bag model, the authors of Ref.~\cite{DiClemente:2022ktz} {studied one SD configuration sitting on the right of the minimum-mass SD in their M-R diagram. They found it is unstable for rapid conversions whereas it becomes stable} if one fixes the total quark baryon number {while varying the transitional pressure (which they argue is equivalent to the slow conversion case)}. {Then, they generalized this particular result to all their obtained SD families having different transitional densities and quark baryon numbers. Nevertheless, we believe that their stable results should be understood carefully since usual radial-oscillation studies consider the transition pressure as constant when imposing rapid and slow boundary conditions. Any modification of this condition will mean distinct physical situations which, as can be seen in Fig. 1 of Ref.~\cite{DiClemente:2022ktz}, even modifies their SD M-R diagram, something unexpected from microphysical conversions. In this sense, despite being interesting results by themselves and which deserve further detailed studies, we will focus in the traditional approach of Refs.~\cite{Haensel:1989wax,Pereira:2017rmp}, in particular, for slow-conversion stars. Thus, all the aforementioned reasons} motivate us to perform a more detailed {and systematic stability} analysis in order to verify if all {the past} conclusions are kept or might change when {the whole families of SDs are studied,} varying also the values of the transitional densities and using another quark model for the ultra-dense core.

In this work, we revisit the question of the stability of strange dwarfs and strange planets\footnote{{By `strange planets' we mean very light planetary-like strange-matter stars whose masses lie in order of the mass of Jupiter ($\sim 10^{-3}~M_\odot$), allowed to exist by the SQM hypothesis.}} by carefully analyzing the behavior of the fundamental mode eigenfrequencies which decide what hybrid stellar configurations are stable or not for the extreme cases of slow and rapid microscopic phase conversions{, both assuming a constant transition pressure}.
We will perform a systematic approach by using three values for the transitional density in order to probe the dependence of the positiveness/negativeness of the zero mode on this free parameter.
Finally, the hadronic crust will be modeled by a fit formula of the original (tabulated) BPS EoS and for the quark phase we will use the MIT bag model as well as an EoS from cold and dense perturbative QCD.

As we will see, the rapid-conversion eigenfrequency of the fundamental mode turns out to be the so-called {\it reaction mode}~\cite{Haensel:1989wax,Pereira:2017rmp}, which generically depends on the energy density jump between phases as well as the ratio between the transition pressure and transition density.
In this sense, it is not obvious that this reaction mode is important in all situations.
We note that this reaction mode has not been properly analyzed in any of the aforementioned works on SDs where large jumps are present at very low critical pressures having also large ratios between critical pressure and transition densities.
In fact, for SDs the reaction mode coincides with the lowest mode, thus being decisive on their stability.

This paper is organized as follows.
In Sec.~\ref{sec:2}, we give a brief summary of the equations of state (EoSs) to be used for the SQM core and for the hadronic mantle.
Besides, some details about the reaction mode in the general-relativistic limit of the Gondek's radial oscillation equations are given. Sec.~\ref{sec:3} is devoted to present our results for strange-dwarfs {and strange planets} analyzing the stability of the stellar configurations for rapid and slow conversions.
Moreover, we {further} investigate the robustness of our results by also using an EoS from cold and dense perturbative QCD to model the SQM core.
Finally, Sec.~\ref{sec:5} discusses and summarizes our main findings for which we also propose possible outlooks. {Along this work we use natural units ($G =\hbar= c = 1$) unless stated otherwise.}

\section{Methodology}
\label{sec:2}
In order to analyze the dynamical stability of strange dwarfs and planets one has to specify the EoSs entering as inputs when modelling their interiors as well as the formalism of infinitesimal radial oscillations when having non-trivial boundary conditions at the interface between phases. The microscopic interior of cold strange dwarfs is described by a matching between an equation of state for dilute nuclear matter to a strange quark matter one through a first-order-like phase transition~\cite{Glendenning:2000} (see, e.g., Refs.~\cite{Glendenning:1994sp,Glendenning:1994zb} for more details).
In this work, we use an analytic fit \cite{Rau:2023} for the BPS model \cite{Baym:1971pw} for the hadronic crust at low baryon densities.
On the other hand, SQM will be described by the massless MIT bag model with $B^{1/4} = {145}{\rm MeV}$ which satisfies $\epsilon/n_{B} < {930}{\rm MeV}$ at zero pressure, being $\epsilon$ and $n_{B}$ the energy density and baryon number density, respectively.
Recall that within this particular model, any $B^{1/4}$ in the range $145~{\rm MeV} \leq B^{1/4} \leq 162.8~{\rm MeV}$ leads to absolutely stable strange quark matter~\cite{Farhi:1984qu}.
Moreover, for the transitional densities, $\epsilon_{\text{t}}$, we use the same values of Refs.~\cite{Glendenning:1994sp,Glendenning:1994zb,Alford:2017vca,DiClemente:2022ktz}, i.e. $10^{9},~10^{10}$ and ${4\times10^{11}}{\rm g/cm^3}$, being the last density\footnote{{These values include implicitly the limit $\epsilon^{\rm max}_{\rm t}= {8.3\times10^{10}}{\rm g/cm^3}$ \cite{Huang:1997b} to ensure mechanical stability.}} associated to the neutron-drip point~\cite{Glendenning:2000}.
%
\begin{figure*}[!t]
  \vspace{-.54cm}
  \hbox{\hspace{-0.5cm}\includegraphics[width=.56\textwidth]{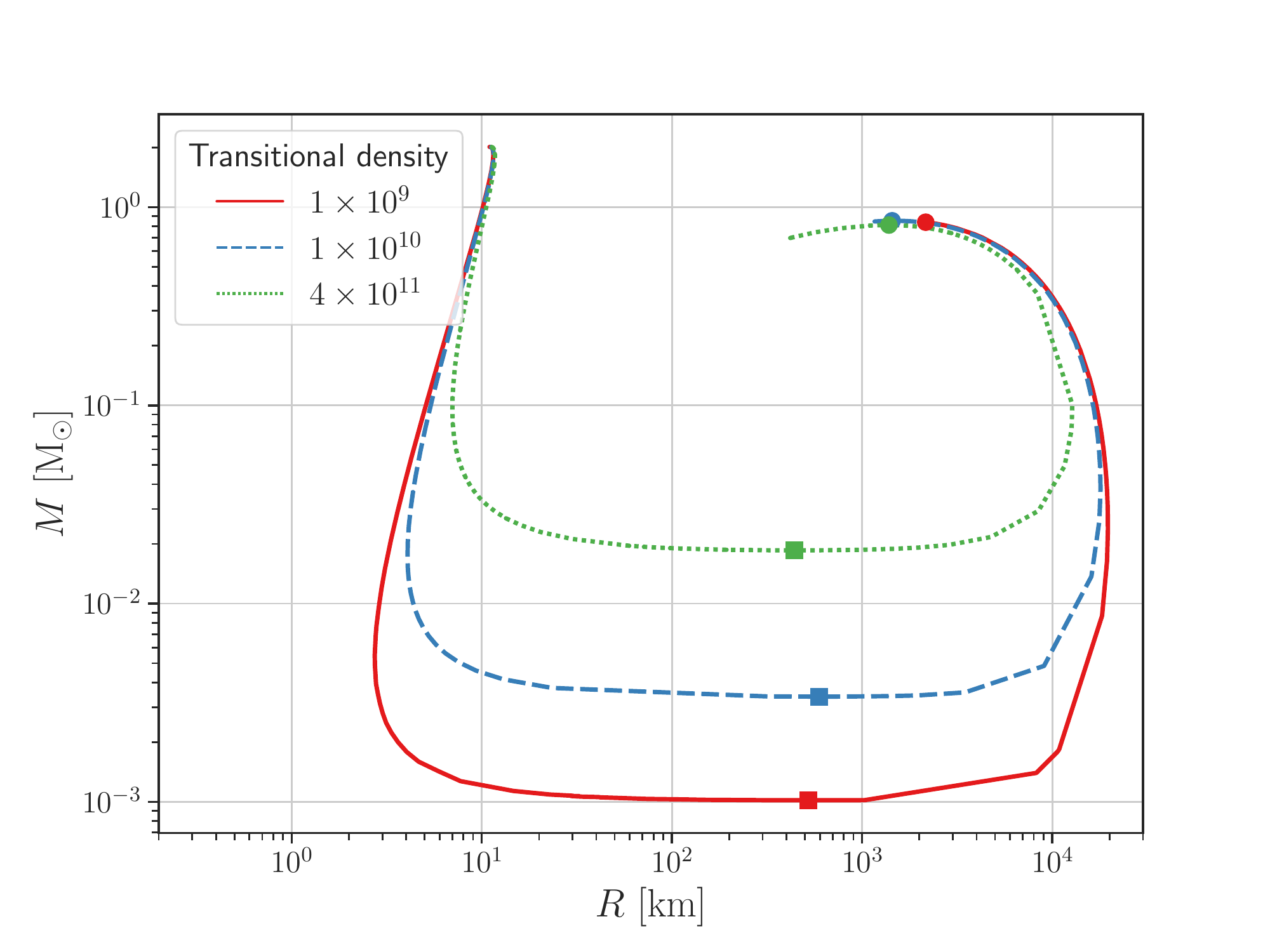}
    \hspace{-0.75cm}\includegraphics[width=.56\textwidth]{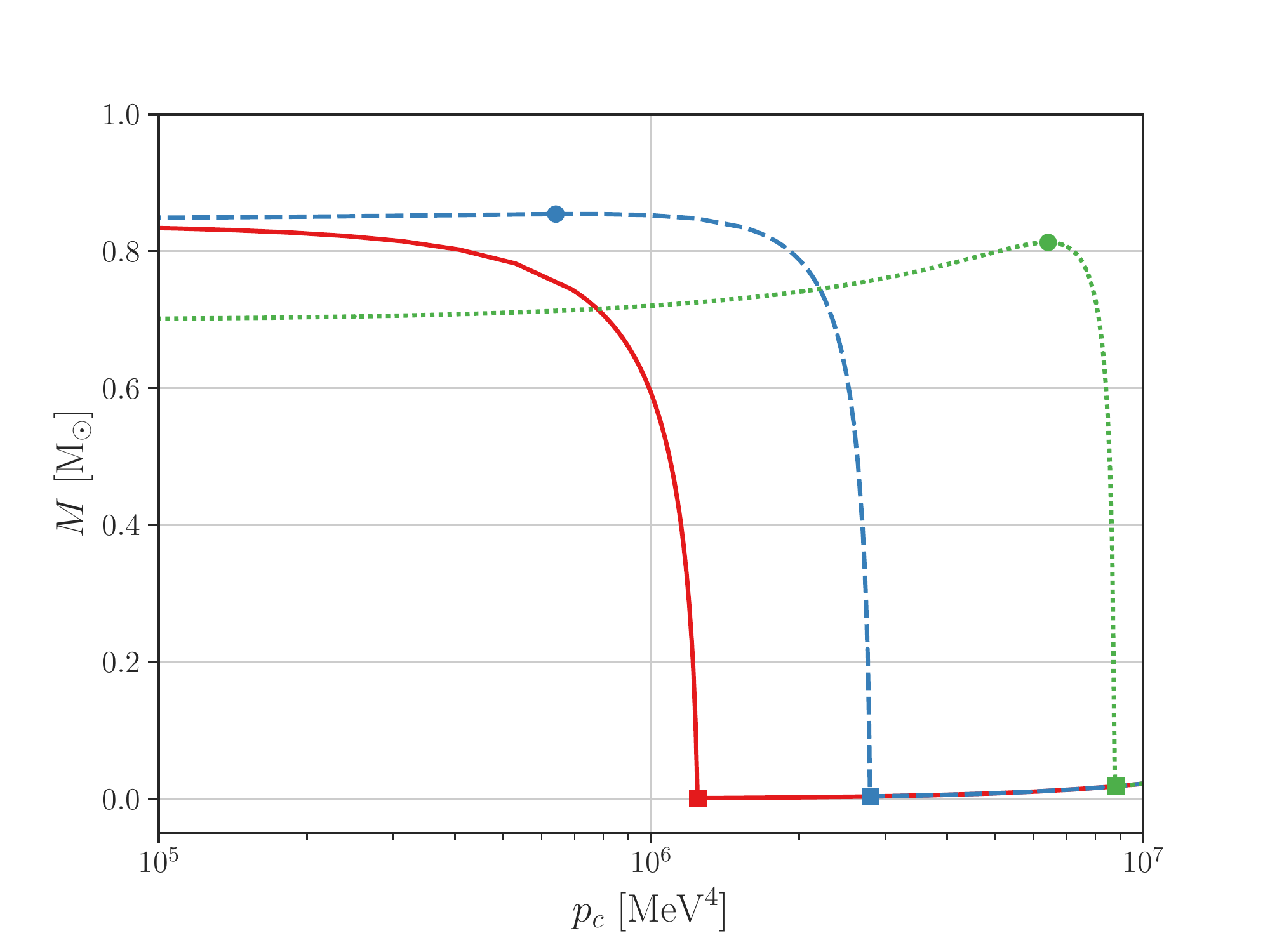}}
  \caption{\label{fig:structureMIT} {{\it Left panel:} Mass-radius diagram indicating the behavior of strange-dwarfs, {strange} planets and usual compact strange stars for different {crust-core} transition densities (in ${\rm g\,cm^{-3}}$) {and using the fit formula for the BPS EoS and MIT bag model}. {For clarity, we are considering strange dwarfs to be between the tip of the curve down to the minimal mass (squares), which partially includes strange planets (where $M \leq 9\times 10^{-2}$ M$_\odot$). At the left of the squares, for $M \geq 0.1$ M$_\odot$ we have the usual strange stars with a thin nuclear matter crust.} {\it Right panel:} {Corresponding} mass-central pressure diagram for strange dwarfs.} The full circles (squares) represent the maximum (minimum) mass configurations for each family of SDs differentiated by a single {crust-core} transitional density value.}
\end{figure*}

In {order to} probe the robustness of our results {for a different quark model at high densities}, we will also consider the state-of-the-art EoS obtained using results of cold and dense perturbative QCD (pQCD) of Ref.~\cite{Kurkela:2009gj} up to next-to-next-to-leading order in the strong coupling $\alpha_{s}$ for a gas consisting of {massless} up and down {plus massive} strange quarks.
Their three-loop calculation gives a pressure parametrized only by the so-called (dimensionless) renormalization scale $X$ which is allowed to vary between 1 and 4 \cite{Kurkela:2009gj}.
In order to get fully renormalization-group invariant results, they also allowed the strong coupling and strange quark mass to be dependent on $X$.
Notice that this scale $X$ is a nonphysical artifact originated when applying dimensional-regularization techniques.
The main astrophysical implications of a pQCD EoS for strange stars have been discussed e.g. in Refs.~\cite{Kurkela:2009gj,Jimenez:2019iuc,Goncalves:2020joq}.
In the next section, we will use the original (involved) numerical results of Ref.~\cite{Kurkela:2009gj}, dubbed pQCD $[X]$, satisfying $\beta-$equilibrium and local electric charge neutrality.
For this  pQCD $[X]$ EoS we restrict ourselves to values of $X$ satisfying the SQM hypothesis, i.e. $X\in[3,4]$.

On the other hand, the stability of strange dwarfs and planets against radial oscillations will be investigated using the Gondek's first-order formalism~\cite{Gondek:1997fd}, which we summarize in what follows.
This formalism is based on the analysis of the variation of pressure $\Delta{p}$ and the radial displacement function $\Delta{r}$, which allows to define the quantity  $\xi=\Delta{r}/r$, both at a given radial coordinate `$r$' and time `$t$' which is factorized as a harmonic contribution.
Such quantities are the Lagrangian variables associated to the radial disturbances of fluid elements inside the general-relativistic star found initially in hydrostatic equilibrium without still knowing if they represent stable or unstable stellar configurations.
Within general relativity\footnote{Although white dwarfs and strange dwarfs (planets) are not fully relativistic objects as a whole since they have very small compactness, $M/R \ll 1$, it is not clear for us if it is completely correct to use the Newtonian approximation for their structure and stability since the SQM core radius, $R_{\rm SQM}$, is very compact compared to the observable radius, $R$, i.e. $R_{\rm SQM} \ll R$, having perhaps relativistic corrections when obtaining their stellar properties. Thus, we study the problem in full relativistic gravity in order to not overlook any of these possible corrections.}, we first solve the Tolman-Oppenheimer-Volkov (TOV) equations to obtain energy density, pressure, and metric function profiles for a given central energy density.
We then insert these profiles into the radial oscillation equations satisfying intuitive boundary conditions (see Ref.~\cite{Gondek:1997fd} for further details) which must give us frequencies $\omega^{2}_{n}$ satisfying $\omega^{2}_{n=0} < \omega^{2}_{n=1} < \omega^{2}_{n=2} < \cdots$ where we recognize $\omega^{2}_{n=0}$ as the (squared) fundamental mode eigenfrequency which help us to analyze the dynamical stability of a given relativistic star.
In practice, if and only if $\omega^{2}_{n=0}$ is (zero) positive, then the  star is said to be (maximally) stable~\cite{Chandrasekhar:1964zza,Gondek:1997fd,DiClemente:2020szl}. {See the Appendix for more technical details on the numerical resolution of the TOV and Gondek's radial oscillation equations with respective (rapid and slow) boundary conditions.}

\begin{figure*}[t]
  \vspace{-0.54cm}
  \hbox{\hspace{-0.5cm}\includegraphics[width=.56\textwidth]{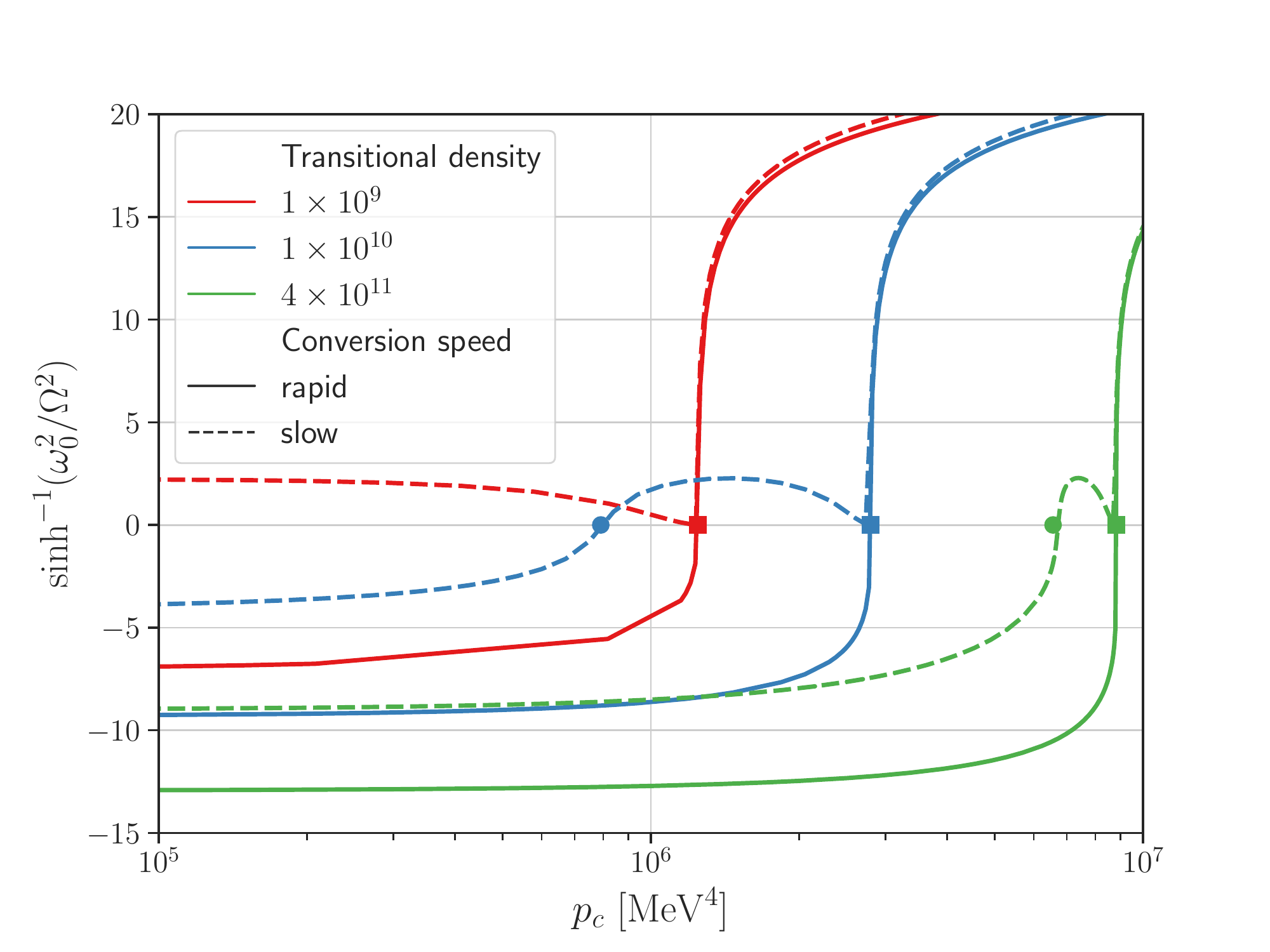}
    \hspace{-0.75cm}\includegraphics[width=.56\textwidth]{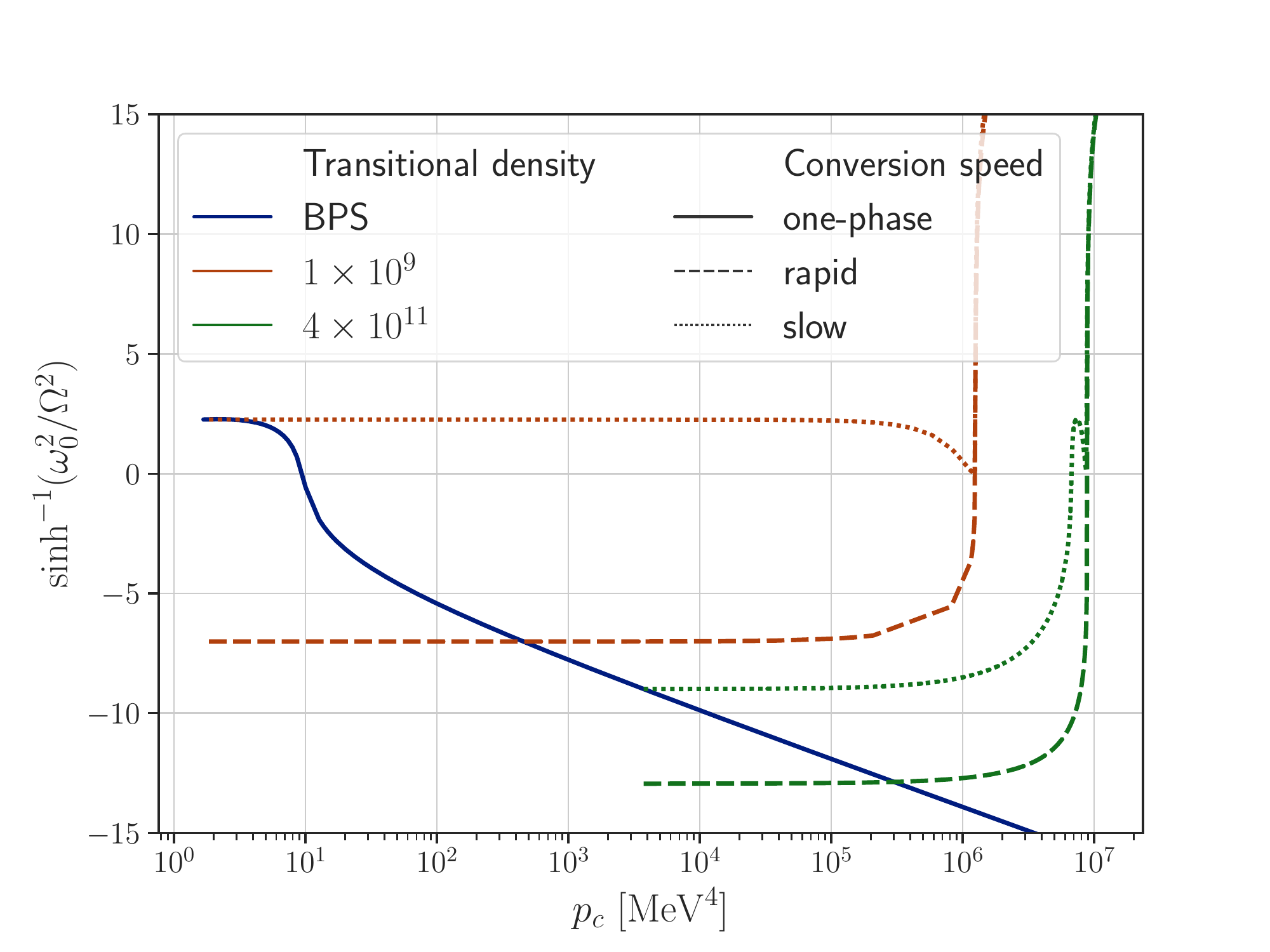}}
  \caption{\label{fig:stabilityMIT} {\it Left panel:} The dependence of the fundamental eigenfrequencies $\omega_0^2$ (through an auxiliary function) with central pressure for each conversion speed and the same {crust-core} transitional densities of Fig.~\ref{fig:structureMIT} in the strange dwarfs and strange planets branch. {\it Right panel:} $\omega^2_0$ for the pure BPS (one-phase, blue solid curve) and SDs with transitional densities of $10^9$ (orange curves) and $4\times 10^{11} {\rm g\,cm^{-3}}$ (green curves) considering a wider central pressure range and, in the SDs case, both conversion speeds. One can see that $\omega^2_{0}$--rapid does not converge to the one-phase (BPS) results. {Although not shown, we have checked that $\omega^2_1$--rapid tends to $\omega^2_0$--one-phase, as expected from the results of Ref.~\cite{Haensel:1989wax}.} As proven in this panel, the rapid cases are in fact the corresponding reaction modes.}
\end{figure*}

For an EoS having a continuous behavior in the energy density, instead of solving the oscillation equations one can use the practical rule of $\partial M / \partial \epsilon_{c} \geq 0$ to discriminate stable from unstable branches of relativistic stars~\cite{Glendenning:2000}.
Nevertheless, if the system is characterized by an EoS that has a discontinuity characteristic of first-order transitions, in particular, by a Maxwell construction\footnote{The smooth Gibbs-Glendenning construction enforcing global electric charge neutrality was studied in, e.g. Refs. \cite{Sahu:2001iv,Gupta:2002fk}.}, then one has to add further boundary conditions {on the radial-oscillation variables} at the phase boundary between SQM and dilute nuclear matter.
This more complex situation was studied in a Newtonian framework in Ref.~\cite{Haensel:1989wax} whereas a general-relativistic generalization was only given some years ago in Ref.~\cite{Pereira:2017rmp}.
In both studies, the extreme cases of rapid and slow phase conversions were explored.
By rapid conversions we refer to the physical situation where the radial oscillation period is {very} large compared to the microscopic timescale leading to the conversion of nuclear matter into SQM (and vice-versa) at the phase-splitting boundary.
On the other hand, slow conversions are characterized by a radial oscillation period that is very small compared to the microscopic timescale~\cite{Haensel:1989wax,Pereira:2017rmp}.

Furthermore, it should be emphasized that for rapid conversions a new mode of oscillation is present, the so-called {\it reaction mode}\footnote{Strictly speaking, this mode has no new mathematical properties, as for quasi-normal modes \cite{Kokkotas:1999bd}, being an ordinary radial mode although highly sensitive to the discontinuity of the rapid junction condition. Besides, it should not be confused with the non-radial reaction modes related to rotational instabilities \cite{Kokkotas:1999bd}.}, already found in Newtonian gravity~\cite{Haensel:1989wax} as well as in general relativity~\cite{Pereira:2017rmp}, which is related to the energy-density discontinuity between phases $\eta\equiv \epsilon^{(-)}_{\rm SQM}/\epsilon^{(+)}_{\rm NM}$ {and the ratio $p_{t}/\epsilon^{(+)}_{\rm NM}$, where $\epsilon^{(+)}_{\rm NM}$ is the maximum nuclear matter (NM) energy density before reaching the discontinuity, $\epsilon^{(-)}_{\rm SQM}$ is the minimum SQM energy density after the discontinuity, and `$p_{t}$' is the transition pressure.}

As noted in Ref.~\cite{Pereira:2017rmp}, the reaction mode (squared) frequency can be estimated as $\omega^{2}_{R}\sim (3[1+p_{t}/\epsilon^{(+)}_{\rm NM}]-2\eta)/(\eta-1)$.
From this relation one can see that this reaction mode can be any excited state, in particular, {its frequency can be} very large {numerically (compared to the usual fundamental and first excited frequencies)} when $\eta \to 1$, i.e. the {size of the} SQM core tends to be very small compared to the {size of the} nuclear matter mantle.
  {Besides, even if this last case applies or not, $\omega^{2}_{R}$ could still be real and a low mode if $p_{t}/\epsilon^{(+)}_{\rm NM}$ is large.}
Thus, the above estimate of $\omega^{2}_{R}$ allow us to infer if the hybrid stars will have real or complex `$\omega_{R}$' {depending on the situation under study}.
For instance, {it was proven \cite{Pereira:2017rmp} that} hybrid compact stars will not have reaction modes when $3[1+p_{t}/\epsilon^{(+)}_{\rm NM}]>2\eta$ is not satisfied or $\eta$ is very large.
Finally, it is worth to mention that Ref.~\cite{Pereira:2017rmp,Zdunik:1987xaa} made these estimates on the assumption that hybrid stars satisfy $\partial M / \partial p_{c} \geq 0$, which {is not always the case.}

Notice that although in Refs.~\cite{Haensel:1989wax,Pereira:2017rmp} this situation was theoretically analyzed, it was not clear which physical system might embody such a condition.
As we will see, strange dwarfs {and planets} are perfect examples where this circumstance occurs. In practice, these reaction modes can be distinguished by analyzing their behavior when the squared eigenfrequencies for the hybrid star, $\omega^{2}_{\rm hyb, n}$, does not tend to the corresponding one-phase {(in our case the BPS model)} values, $\omega^{2}_{\rm one-phase, n}$, as we reduce the quark matter core to zero, for a given mode `$n$' mode.
Of course, if this behavior for the reaction modes is found for the fundamental mode ($n=0$) then the whole dynamical stability will be determined exclusively by how one {tunes the values of} `$\eta$' {and `$p_{t}/\epsilon^{(+)}_{\rm NM}$' from physical arguments characterizing a} family of hybrid neutron or strange dwarfs (planets).

\begin{figure*}[t]
  \vspace{-0.54cm}
  \hbox{\hspace{-0.5cm}\includegraphics[width=.56\textwidth]{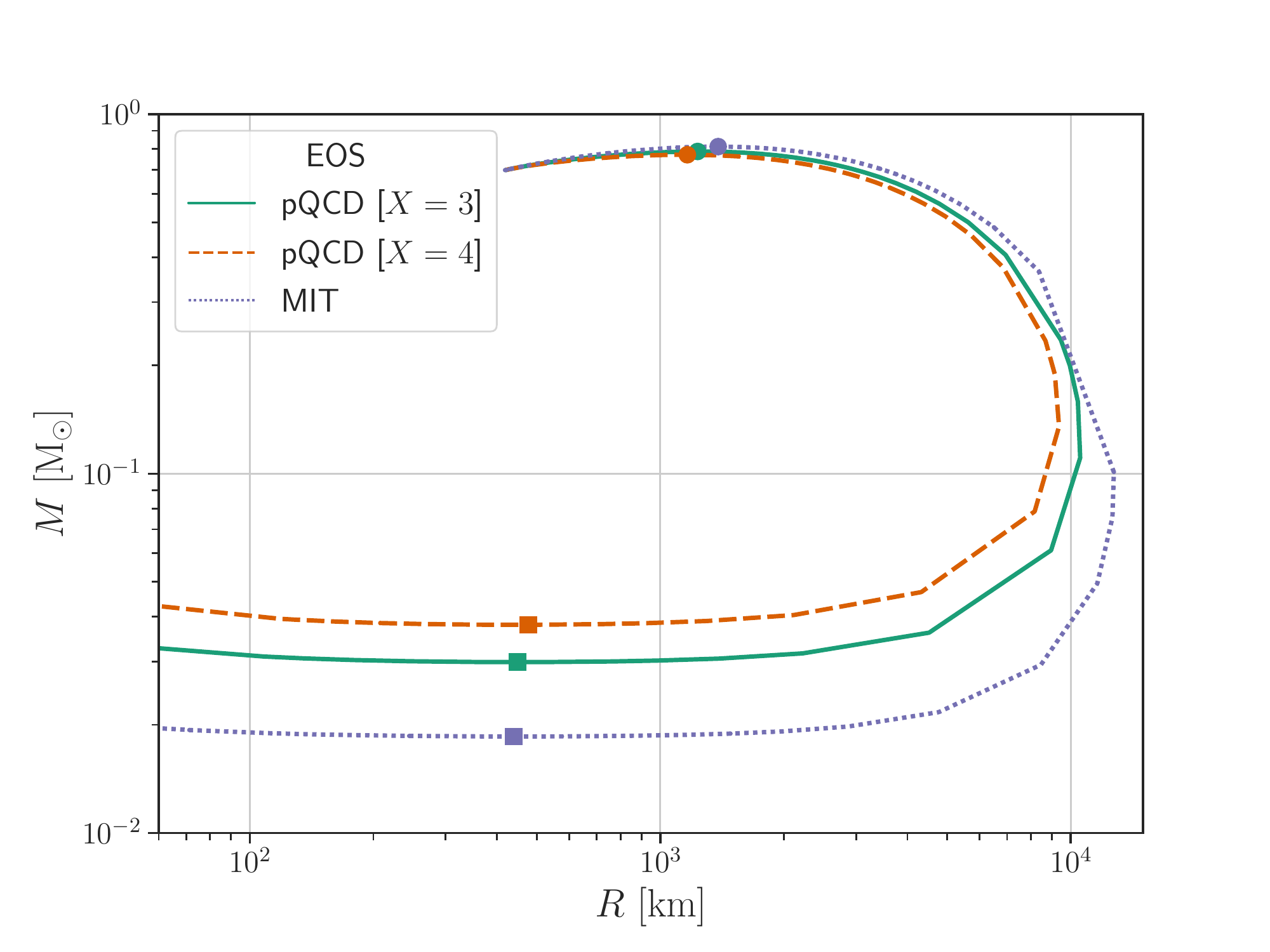}
    \hspace{-0.75cm}\includegraphics[width=.56\textwidth]{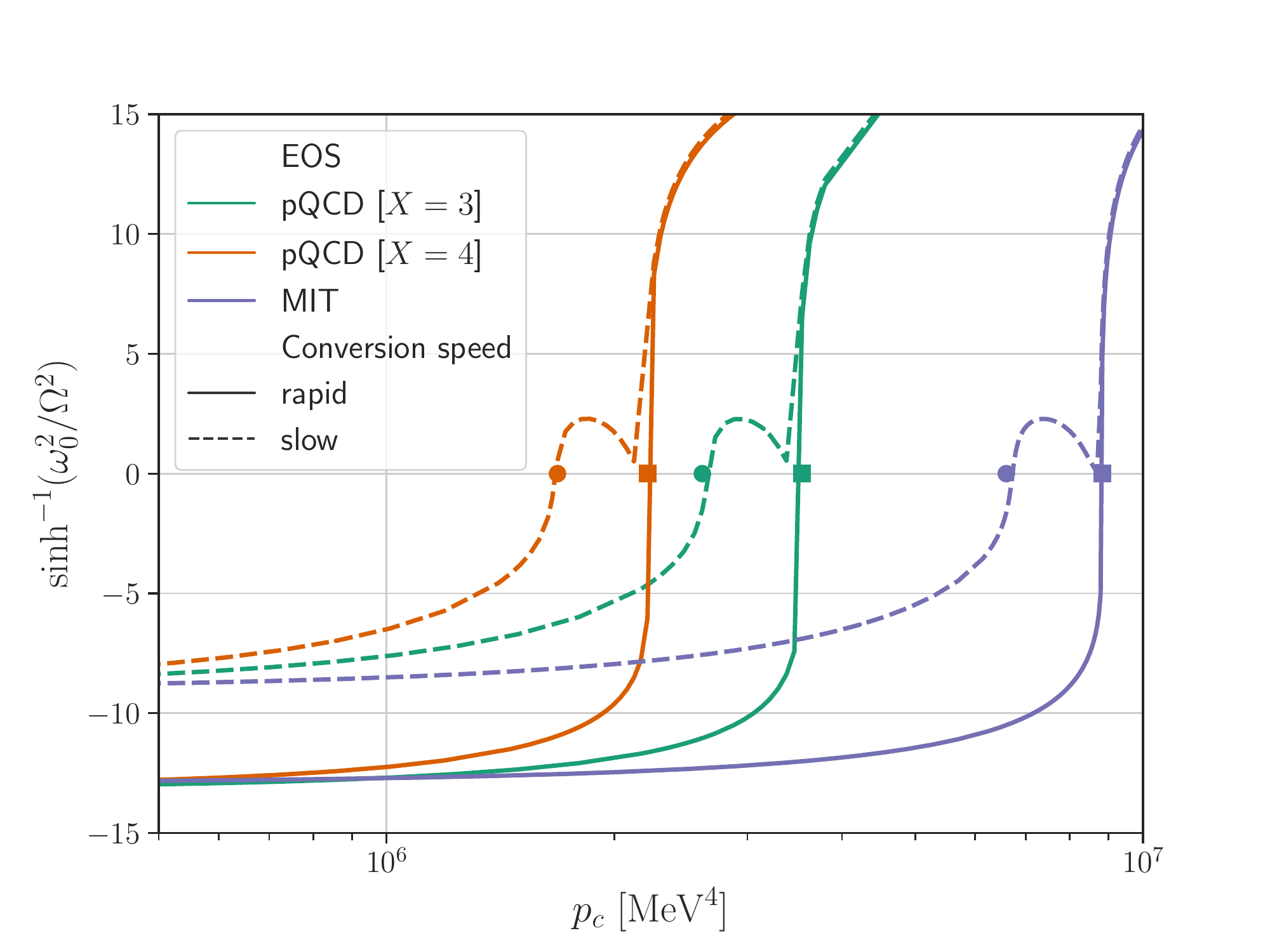}}
  \caption{\label{fig:pQCD} {\it Left panel:} Mass-radius relation for SDs {(planets) obtained using the fitted BPS formula} with the pQCD EoS for quark matter for the extreme $X$ in the pQCD EoS producing stable SQM, in the particular case of the neutron drip density as the {crust-core transition} density. {\it Right panel:} {Corresponding behavior of} $\omega^2_0$ {(through the auxiliary function)} as a function of central pressure. {Again, the zero mode is the reaction mode which is unstable in the strange dwarf (planet) branch.}}
\end{figure*}
%

\section{Results}
\label{sec:3}

In this section, we will display our numerical results for the structure and stability of strange dwarfs (planets).
  {Along this section,} their crust will be modelled {only} by the BPS \cite{Baym:1971pw,Glendenning:2000} equation of state but now adjusted to be a smooth analytic formula \cite{Rau:2023}.
{On the other hand, for the first part of this section we use the MIT bag model for the quark core of these SDs}, whereas in the second {part} we change to the pQCD EoS to explore {if QCD corrections affect their stability or not}.
It should be noted that at the transition point our strange-dwarf EoSs have a discontinuous derivative, although they are continuous by parts.
This agrees with Ref. \cite{Glendenning:1994zb} but {differs from the approach of} Ref.~\cite{Alford:2017vca}, {not only replacing the original sharp transition by a smooth crossover, but also excluding the consistent imposition of the rapid junction conditions when solving the radial oscillation equations.}
We stress that the main objective of this section is to fill the gap in relation to past works, where somewhat incomplete results for the dynamical stability of these stellar objects was presented.
In particular, we aim to probe, strengthen and extent the findings of Refs.~\cite{Glendenning:1994sp,Glendenning:1994zb,Alford:2017vca,DiClemente:2022ktz}.

\subsection{BPS + MIT strange dwarfs}
In Fig.~\ref{fig:structureMIT}, we present our results for the hydrostatic structure of strange stars, strange dwarfs and strange planets within the massless MIT bag model EoS for different values of {the crust-core} transition densities, i.e. $10^9,~10^{10}$ and ${4\times10^{11}}{\rm g\,cm^{-3}}$.
Throughout this section the filled circles and squares denote the maximum and minimum mass SDs, respectively, with a corresponding color for each {crust-core} transitional density.
In the left panel, we present the mass-radius relation for the hybrid family of stars composed by the BPS+MIT EoSs, which does not include pure WDs.
In the $M$-$R$ relation, we are able to identify the branch of strange dwarfs and strange planets with masses in the range $\sim {0.8} \to {0.1}{M_{\odot}}$ and $\sim 9\times10^{-2} \to 10^{-3} {M_{\odot}}$, respectively, between the beginning of the curve (left of the circles) and the squares.
We also show the mass-central pressure diagram in the right panel, where one can see that the SD branch begins at the transition density/pressure (not shown, because it occurs for even smaller central pressures), passes through the points of maximum mass (circles) and ends with configurations of minimum mass (squares), where the compact strange-star branch begins.

We now pass to investigate the stability of SDs through a radial oscillations' analysis for both rapid and slow conversions.
Following Ref.~\cite{Alford:2017vca}, we display our findings using the auxiliary function $\sinh^{-1}(\omega_n^2/\Omega^2)$ (which preserves the sign of $\omega_n^2$) as a function of central pressure $p_{c}$ and $\Omega ={6.6\times10^{-22}}{\rm MeV}$ being an appropriate constant.
In Fig.~\ref{fig:stabilityMIT} (left panel), we present our results for the stability of SDs with different transitional densities.
Considering rapid conversions (solid curves), we have that all SD configurations at the left of the minimum mass are unstable i.e. $\omega_0^2 < 0$, which -- as expected from the results of Ref.~\cite{Pereira:2017rmp} -- give the same results if one applies the usual stability criterion $\partial M/\partial p_c > 0$ (see the right panel of Fig.~\ref{fig:structureMIT}).
On the other hand, in the slow conversion case they are stable against radial oscillations.

Both Refs.~\cite{Glendenning:1994zb} and~\cite{Alford:2017vca} did not present a clear treatment of the radial oscillations' equations at the phase-splitting surface required to compute the stability of hybrid stars, since both of them did not explicitly impose the extra boundary conditions proposed in Ref.~\cite{Pereira:2017rmp}.
In this sense, the discordance in their results comes from the treatment of the phase transition itself, i.e. while in Ref.~\cite{Glendenning:1994zb} the BPS and the MIT EoSs were combined to form a discontinuous hybrid EoSs, in Ref.~\cite{Alford:2017vca} the same EoSs were combined as a smooth crossover.
Moreover, in Ref.~\cite{Glendenning:1994zb}, the combination of the two EoSs (BPS and MIT EoSs) in a discontinuous way without extra boundary conditions means that they were (implicitly) assuming the continuity of the Lagragian variables which is equivalent to the slow conversion scenario.
In contrast, the treatment of Ref.~\cite{Alford:2017vca} implies they were essentially dealing with a one-phase star, in the sense that there were no discontinuities in the EoS which means that the usual stability criterion still holds, as is the case of rapid conversions in the formalism of Ref.~\cite{Pereira:2017rmp}.
Thus, the left panel of Fig.~\ref{fig:stabilityMIT} qualitatively contains the results of both Refs.~\cite{Glendenning:1994zb} and~\cite{Alford:2017vca} and their divergences which can now be explained through the different conversion speeds, in agreement with the brief discussion of Ref.~\cite{DiClemente:2022ktz}.

Another important feature that helps to further explain the aforementioned discordances between Refs.~\cite{Glendenning:1994zb} and~\cite{Alford:2017vca} is the reaction mode.
This mode appears in the rapid conversion scenario and it is defined as the mode whose frequency does not tend to its one-phase counterpart when $p_c \rightarrow p_t$ or equivalently, in our case, $R_{\rm QM} \rightarrow 0$.
As explained in Ref.~\cite{Haensel:1989wax}, when the core is much smaller than the outer layers of the star one has that the reaction mode is the fundamental one, being decisive in the determination of the star's stability.
In the right panel of Fig.~\ref{fig:stabilityMIT}, we present $\omega_0^2$ as a function of central pressure -- considering a large interval of $p_c$ -- for the pure BPS (blue solid curve) but also for the BPS + MIT hybrid configurations with {crust-core} transition densities of $10^9$ (orange curves) and $4\times 10^{11} {\rm g\,cm^{-3}}$ (green curves) in the rapid (dashed) and slow (dotted) scenarios.
From this figure, one can clearly see that the fundamental frequency in the slow conversion scenario for both transitional densities coincide with the one-phase BPS at the beginning of the respective dotted curves, which is the region where $p_c \rightarrow p_t$.
Additionally, one can notice that the reaction mode have negative (squared) zero-mode frequencies over a large region of central pressures resulting in unstable SDs, specially for $\epsilon_t = 10^9 {\rm g\, cm^{-3}}$ where its slow frequencies and the BPS are positive.
As already discussed, this reaction mode was not analyzed in Refs.~\cite{Glendenning:1994zb} and~\cite{Alford:2017vca} since the former work was implicitly dealing with slow conversions while the last work smoothed the phase transition, thus leading to an artificial continuity between the pure BPS frequency and the fundamental one of SDs.
Besides, although it seems that the reaction mode was also found in Ref.~\cite{DiClemente:2022ktz}, it was not properly identified/characterized since the authors only calculated the frequency of the fundamental mode (of the rapid case) for a single configuration near the minimum mass.
We believe that all these technical issues posed this SD stability puzzle for which we proved that they are only stable in the slow scenario.

For completeness, some comments must be made about the solutions of the Lagrangian variables, in particular, $\xi$.
Generically, when going from the core to the neighborhood of the transition radius of the crust ($\sim {4}{\rm km}$) the adiabatic index will decrease from being very large (almost incompressible quark matter) to be very small (very compressible WD matter).
This will induce regular values of $\xi\sim 1$ at the core but large values of `$\xi$' in the dilute mantle.
One can easily see this by noticing how the adiabatic index enters into the Gondek's oscillation equations.
Of course, depending on the conversion speed one finds a (dis)continuous behavior of `$\xi$' at the transition point.
We have also verified in all our calculations that the obtained `$\xi$' {agree with the corresponding number of nodes, for instance, `$\xi$' with} no nodes are related to the fundamental mode.

\subsection{BPS + pQCD strange dwarfs}
Finally, in order to probe the robustness of our findings for the stability of strange dwarfs (planets), we present some results for the M-R diagram and behavior of $\omega_0^2$ but now using the pQCD EoS and keeping the BPS fit formula for the thick crust {fixing also the transitional density to be the neutron drip one.}
When the pQCD [$X = 4$] EoS is used we find SDs with {slightly} smaller maximum masses while the corresponding strange planets are heavier, all this compared to the {pQCD [$X=3$] and} MIT bag model, as can be seen in Fig.~\ref{fig:pQCD} (left panel).
Besides, {a similar behavior is found for} the case of pQCD [$X = 3$] {when compared to} the MIT bag model.
Physically, all this happens because SDs around the maximum mass have a smaller energy density jump for the pQCD [$X$] (for increasing $X$) EoS than the bag model. On the other hand, configurations near the minimum mass have larger SQM cores and are described by stiffer EoSs which imply in higher masses for strange planets when pQCD [$X$] (for increasing $X$) is compared to the bag model. In short: maximum masses are dominated by the transition jump whereas masses near the minimum are dominated by a large SQM core. Notice that the same reasoning applies for the discussion at the beginning of this section for the bag model with different transitional densities.
Now, in the right panel of Fig.~\ref{fig:pQCD}, we present the stability analysis of the pQCD SDs having almost the same qualitative behavior as when using the MIT bag model although the pQCD strange dwarfs and planets live in the sector of lower pressures compared to the bag model.
This can be understood by noting that the pQCD EoS is stiffer for increasing $X$ which in strange star branch produces larger maximum masses.
Nonetheless, rapid conversions still rule SDs and planets out whereas in slow conversions strange dwarfs and planets remain stable.
Furthermore, we also proved that the BTM criteria is still true only for rapid conversion SDs.
In summary, the MIT bag model and pQCD EoSs for high densities produce similar results when such low-density transitions occur.

\section{Summary and outlook}
\label{sec:5}

We have revisited the question concerning the dynamical stability of strange dwarfs and planets for which there was still not a clear answer about its in-principle existence in nature.
For this we have performed a detailed analysis of the behavior of the fundamental mode eigenfrequencies obtained after solving the radial oscillation equations within the Gondek's formalism \cite{Gondek:1997fd} but now also including non-trivial boundary conditions for their Lagrangian displacements when the discontinuity is found at the interior of these strange objects~\cite{Haensel:1989wax,Pereira:2017rmp} and assuming that the related microphysics follows the extreme situations of rapid and slow phase conversion dynamics.
For slow conversions, our results qualitatively agree with what was presented in Ref.~\cite{Glendenning:1994zb}. However, in the rapid scenario the fundamental mode becomes the reaction mode {but providing only unstable eigenfrequencies, thus} ruling out any stable strange-dwarf configurations, as also suggested by the BTM criteria.
Moreover, our results on the stability of SDs and planets appear to be robust when using an EoS obtained from cold and dense perturbative QCD.

Finally, an immediate generalization of our studies would be to explore other junction conditions, e.g. the ones explored in Refs.~\cite{Karlovini:2003xi,Rau:2022ofy}.
{For instance, the intermediate (timescale) case between rapid and slow conversions might allows us to probe its macroscopic effects even without having detailed microscopic knowledge of the physics occurring at the crust-core transition point. Of course, additional studies must be done elsewhere to clarify this issue in strange dwarfs.}
Besides, it seems interesting to study the gravitational waves coming from the merger of NS and white dwarfs \cite{Paschalidis:2011ez} but now allowing the dwarf star to have a SQM core as well as a strange quark star merging with a strange dwarf.
For physics beyond the standard model, it might be possible to find sizeable variations of our findings within modified theories of gravity, as already done for white dwarfs \cite{Astashenok:2022kfj}.

\begin{acknowledgments}
  We thank the anonymous referees for constructive comments which have improved our work. We thank Steve Harris and Peter Rau for useful comments on the first version of this manuscript as well as very useful discussions. This work was partially supported by INCT-FNA (Process No. 464898/2014-5). V.P.G. and L.L. acknowledges support from CNPq, CAPES (Finance Code 001), and FAPERGS. V.P.G. was partially supported by the CAS President's International Fellowship Initiative (Grant No.  2021VMA0019). J.C.J. acknowledges support from FAPESP (Processes No. 2020/07791-7 and No. 2018/24720-6).
\end{acknowledgments}
\section*{APPENDIX: \\DYNAMICAL STABILITY OF \\ RELATIVISTIC STARS}

In principle, studies concerning the dynamical stability of relativistic stars must be performed within the ADM-formulation of numerical relativity, also known as (3+1)-decomposition of the Einstein equations. Interestingly, it can be proven (see, e.g. Chap. 3 of Ref. \cite{Ruoff:2000nj}) that if one applies time-dependent perturbation theory to these set of complicated equations assuming disturbances with small amplitude around a certain hydrostatic solution, one obtains the same Sturm-Liouville problem as originally derived by Chandrasekhar in 1964. In fact, most general-relativistic hydrodynamic codes test the reliability of their findings by comparing them against well-known perturbative results \cite{Font:2001ew} for radial oscillation frequencies. This approach has also been taken for boson stars where the eigenvalue problem becomes much more complicated than a hydrodynamic study through a Fourier transform but which in the end serves to verify the agreement between both independent methods \cite{Kain:2021rmk}.

For this work, we solved one of the first-order equivalent forms of the Sturm-Liouville problem, i.e. the Gondek's set of equations, being more amenable numerically. First, one needs to solve the TOV equations~\cite{Glendenning:2000} given by
\begin{align}
  \frac{dp}{dr} & = -\frac{\epsilon m}{r^2}\left[1 + \frac{p}{\epsilon}\right]\left[1 + \frac{4\pi r^3 p}{m}\right]\left[1 - \frac{2m}{r}\right]^{-1} \,, \\
  \frac{dm}{dr} & = 4\pi r^2 \epsilon \,, \hspace{2cm} \frac{d\nu}{dr}  = -\frac{1}{p + \epsilon}\frac{dp}{dr}\,,
\end{align}
where $p, \epsilon$, $m$, $\nu$ and $\lambda$ are the pressure, energy density, mass, and metric functions, respectively.
The integration of these equations starts from a central pressure $p(r=0)=p_c$ and ends when $p (r=R) = 0$, thus  extracting the corresponding radius and, consequently, its mass $M = m(r=R)$, which starts at $m(r=0) = 0$ to assure regularity at the star center. Besides, the surface boundary condition on `$\nu$' is $\nu(r=R) = ({1}/{2})\ln(1 - {2M}/{R})$. Now, the solutions of the TOV equations enter as coefficients into the (Gondek's) radial oscillation equations given by
\begin{align}
  \frac{d\xi}{dr}      & = \left(\frac{d\nu}{dr}-\frac{3}{r}\right)\xi - \frac{1}{r}\frac{1}{\Gamma p} \Delta p\,,                                                                 \\
  \frac{d\Delta p}{dr} & = r(p+\epsilon)\left[\omega^2 e^{2(\lambda - \nu)} + \frac{4}{r}\frac{d\nu}{dr} + \left(\frac{d\nu}{dr}\right)^2 - 8\pi e^{2\lambda}p\right]\xi \nonumber \\
                       & - \left(\frac{d\nu}{dr} + 4\pi(p+\epsilon)r e^{2\lambda}\right)\Delta p \,,
\end{align}
where $\xi = \Delta r/r$ and $\Delta p$ are the Lagrangian displacement and the Lagrangian perturbation of the pressure, respectively. Moreover, $\Gamma = (1 + e/p)dp/de$ is the adiabatic index.
The $\omega^2$ are the eigenfrequencies of the star, as discussed in the main text.

Regarding the boundary conditions, we have that to avoid irregularities at the stellar center $\Delta p (r = 0) = -3 \Gamma p \xi(r=0)$.
Since $\xi(r = 0)$ is arbitrary, it is usually taken to be equal to one.
However, as discussed in Ref. \cite{Rau:2023}, for strange dwarfs taking $\xi \ll 1$ improves the computational performance.
The eigenfrequencies $\omega^2$ of the system are the ones that lead to $\Delta p (r=R) = 0$~\cite{Gondek:1997fd}.

In the case of hybrid stars with a strong first-order phase transition, one has to impose extra boundary conditions at the phase-splitting interface for `$\xi$' and `$\Delta p$', depending on the conversion speed~\cite{Pereira:2017rmp}.
For both conversion speeds, one has that $[\Delta p]_-^+ = 0$, meaning that there is no discontinuity in $\Delta p$ at the interface, which comes from the constancy of the pressure during the transition and across the interface.
Therefore, the difference between conversion speeds is computed in the boundary conditions of $\xi$, such that for slow conversions the fluid elements of each phase can be tracked during the oscillation resulting in the continuity of $\xi$, i.e., $[\xi]_-^+ = 0$.
On the other hand, for rapid conversions one has~\cite{Pereira:2017rmp}
\begin{equation}
  \left[\xi - \frac{\Delta p}{r (dp/dr)}\right]_-^+ = 0\,.
\end{equation}
These extra boundary conditions were not included in both Refs.~\cite{Glendenning:1994zb} and~\cite{Alford:2017vca}, and the distinct main conclusion were due to the difference in the treatment of the EOS at the transition.

  {Regarding our numerical procedure, we used a spline interpolation for the tabulated pQCD EoSs since the other EoS are analytical. After that, we applied another spline interpolation for the TOV solutions before being introduced into the Gondek's equations, in particular, care need to be taken when obtaining `$\Gamma$' since it is highly sensitive to numerical derivatives of the already interpolated EoS and metric profiles. Then, the eigenfrequencies are found through a shooting method with the required boundary conditions for `$\xi$' and `$\Delta p$' at the center. The integration is carried out up to the phase-splitting interface where the extra boundary conditions for rapid or slow conversions are applied which gives us the correct values of `$\xi$' and `$\Delta p$' at the other side of the interface. From this point on, the integration continues outwards aiming to match the boundary condition for `$\Delta p$' at the surface. After each integration, the trial value of $\omega^2$ is corrected through a root finding method (in our case, the secant method) until the desired precision on the boundary condition is achieved. These calculations will give the eigenfrequencies $\omega^{2}_{n}$ of the star, being a discrete and numerically ordered set of real numbers (the case `$n=0$' is identified with lowest eigenfrequency value) satisfying appropriately the required rapid and slow junction conditions.}

Finally, a hint for the readers that intend to {reproduce and extend our findings for} the stability of SDs.
In our study we realized that solutions for the TOV equations and the rapid-conversion radial-oscillation equations (either in the Gondek's equations or the Sturm-Liouville problem \cite{Glendenning:2000}), are  sensitive to the numerical treatment of the low density BPS EoS.
This technical issue is important when determining the zero pressure value (which allows us to define the stellar radius) and boundary condition for $\Delta p$ (also vanishing at the SD surface).
{As already announced, we employed the following} fitting formula~\cite{Rau:2023} for the BPS EoS:
\begin{equation}
  \log \epsilon = {a + b\sqrt{1 + c(d + \log p)^2}}\,,
\end{equation}
{where the coefficients are $a = -15.8306,\, b = 11.2974,\, c = 0.00664824$ and $d=16.9824$, with both `$p$' and `$\epsilon$' in MeV$^4$. This fit formula has a smooth behavior at all densities and removes the numerical issues that arise when the interpolation of the tabulated BPS is used. Moreover, the fitting formula offers a better numerical performance. For instance, one can analytically obtain the speed of sound squared}
\begin{equation}
  c_s^2 = \frac{dp}{d\epsilon} = \frac{p\sqrt{1 + c (d + \log p)^2}}{bc(d + \log p)\epsilon},
\end{equation}
{which allows also to calculate analytically the adiabatic index $\Gamma = (1 + \epsilon/p)c_s^2$. Both quantities are determining when solving the radial oscillation equations.}

\end{document}